EUROPEAN LABORATORY FOR PARTICLE PHYSICS

CERN - LHC DIVISION

# Measurements of the Secondary Electron Emission from Rare Gases at 4.2K

*J. BARNARD, I. BOJKO, N. HILLERET*

# MEASUREMENTS OF THE SECONDARY ELECTRON EMISSION FROM RARE GASES AT 4.2K.

John BARNARD, Iouri BOJKO and Noël HILLERET

*LHC-VAC, CERN, 1211 Geneva 23, Switzerland*

Dependence of the secondary electron yield (SEY) from the primary beam incident energy and the coverage has been measured for neon, argon, krypton and xenon condensed on a target at 4.2K. The beam energy ranged between 100eV and 3keV, the maximal applied coverage have made up 12000, 4700, 2500 and 1400 monolayers correspondingly for neon, argon, krypton and xenon. The SEY results for these coverages can be considered as belonging only to investigated gases without influence of the target material. The SEY dependencies versus the primary beam energy for all gases comprise only an ascending part and therefore, the maximal measured SEY values have been obtained for the beam energy of 3keV and have made up 62, 73, 60.5 and 52 for neon, argon, krypton and xenon correspondingly. Values of the first cross-over have made up 21eV for neon, 14eV for argon, 12.5eV for krypton and 10.5eV for xenon. An internal field appearing across a film due to the beam impact can considerably affect the SEY measurements that demanded the beam current to be reduced till $9.0 \bullet 10^{-10}$A. Duration of the beam impact varied between 500μsec and 250 μsec. It was found that reliable SEY measurements can also be taken on a charged surface if the charge was acquired due to beam impact with electrons of higher energy. All SEY measurements for once applied coverage have been carried out for whole range of incident energies from 3keV down to 100eV without renewing the film. Developing of pores inside of a deposited film can significantly increase the SEY as it was observed during warming up the target.

---

## Introduction

The general features of the secondary electron emission (SEE) from metals are well understood and explained in different theories. This was assisted by a large amount of experimental information and also by a fact that the experimental information comprises data for simple chemical elements that is very important for the finding of a reliable SEE theory. This is not the case for insulators. All available until now experimental data for insulators concern mainly with metal compounds representing complex chemical elements but not the simple ones. In this light the secondary electron emission from rare gases which are in fact insulators seems to be an attractive material for creation of the solid SEE theory for insulators. But from another side the SEE data from rare gases still remain until now unavailable for scientific world.

The present article aims to present some experimental data of SEE obtained for neon, argon, krypton and xenon condensed on a target at 4.2K temperature. A description of experimental set-up, method of the SEE measurements and an overview of faced problems will be given.

## Experimental set-up

Taking into account the values of the saturated vapour pressures for the rare gases at different temperatures one can do a conclusion that only the temperatures closed to the ones of the liquid helium can create ultra-high vacuum conditions necessary to pass an electron beam. That is why all SEE measurements for rare gases have been carried out on a set-up having a target where the injected gases are condensed, at 4.2K. The general view of the experimental set-up is shown on Fig.1. The set-up consists mainly of a cryostat with liquid helium, ultra-high vacuum chamber bakeable till $350^0$C, also bakeable injection line equipped with a diaphragm for calculation of throughput of the gas being injected, vacuum gauges, a residual gas analyzer, an electron gun and pumping tools for ultra-high and insulating vacuum.

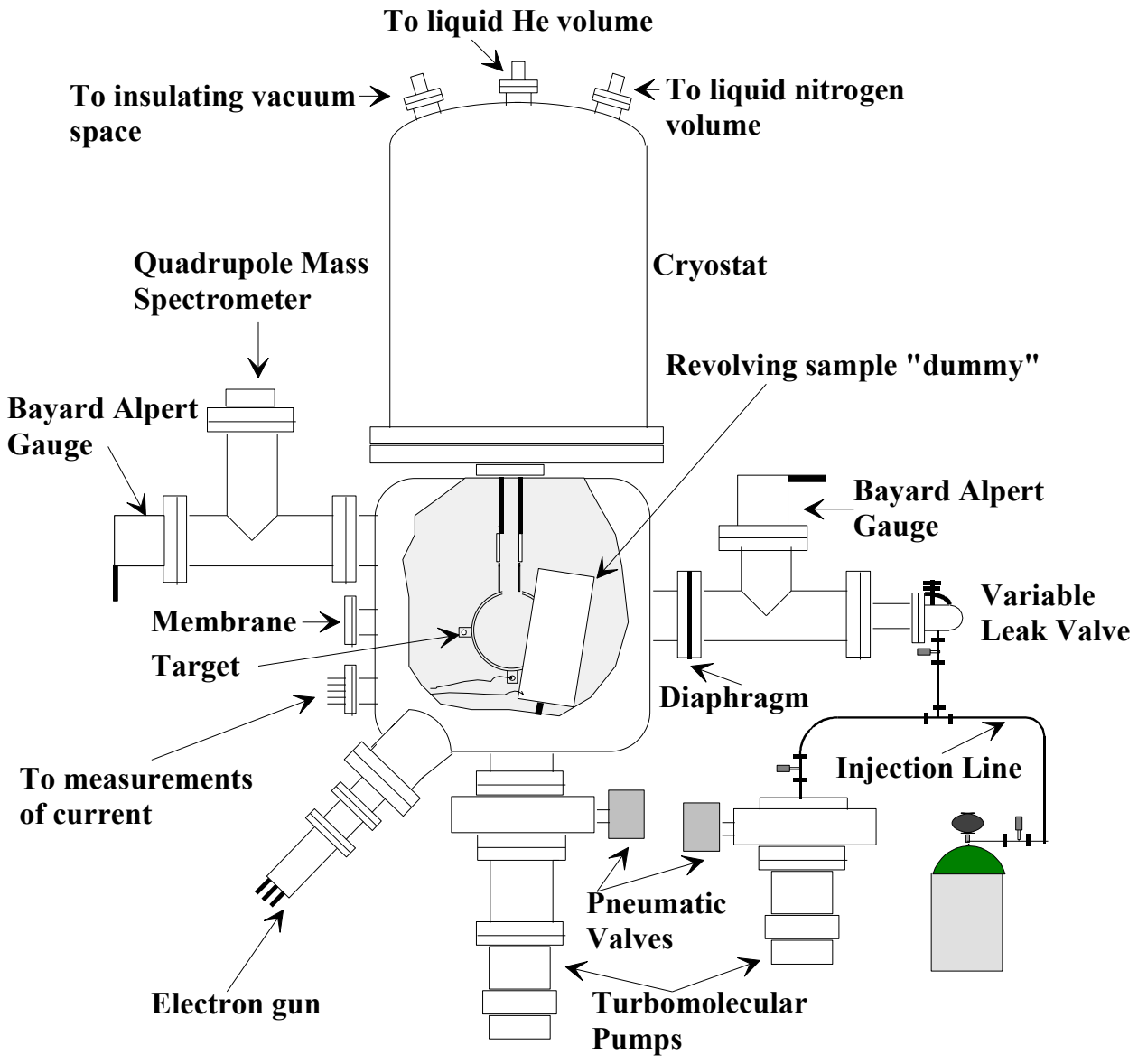

*Fig. 1.General view of the experimental set-up.*

With a hole in the diaphragm being of 9.5mm diameter it provides the conductance for a molecular flow of 8.5l/sec nitrogen equivalent. The throughput of the gas being injected is calculated as product of the conductance to the difference of the pressures before and after diaphragm. The turbopump installed on the vacuum chamber serves to evacuate the system down to an ultimate pressure and it is valved off during gas injection. The vacuum chamber comprises the cold target and a revolving sample

“dummy” made of copper and serving for protection of a condensed film from electron beam during the gun adjustment and also for determination of the beam current. (A rotating drive for the “dummy” is not shown on the Fig.1).

The cryostat represents itself a modified cryogenic pump[1] and its view is shown on Fig.2. Since the present set-up was not equipped with tools allowing direct measurements of the thickness of the film on the target, a question of an exact area of the pumping surface was of major importance. A liquid helium vessel as well as liquid nitrogen screens have been isolated from ultra-high vacuum volume. The target placed in UHV volume and shown on the Fig.3 was connected with the liquid helium vessel via ∅8/6 stainless steel tube. The tube in its UHV part was surrounded with two bellows playing a role of a compensating element and a thermal screen. One bellow was connected to UHV jacket with ambient temperature, another one was connected to the liquid nitrogen screen. Such a solution ensured the keeping a part below the last bellow at liquid helium temperature whilst the bellows had higher temperatures and correspondingly much less pumping speed. The target represents a flattened, hollow cylinder of stainless steel which is mounted onto the liquid helium vessel via a short Pyrex tube providing an electrical isolation. The exterior of the target has been electroplated with copper. The target was of 20mm diameter and of 8mm height. Only the flat plat of the target was exposed to the beam impact.

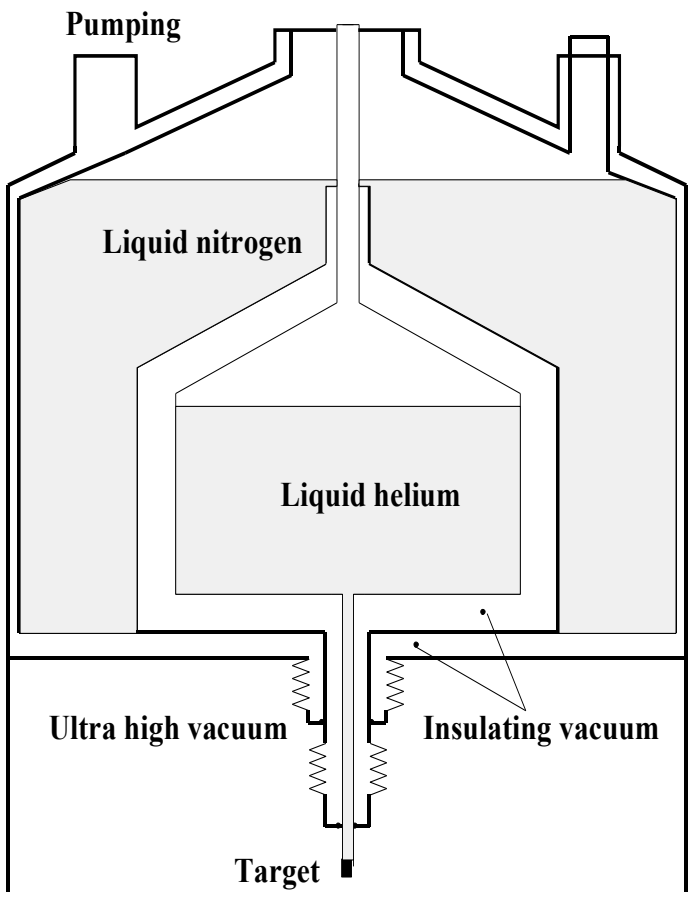

*Fig. 2.Cryostat.*

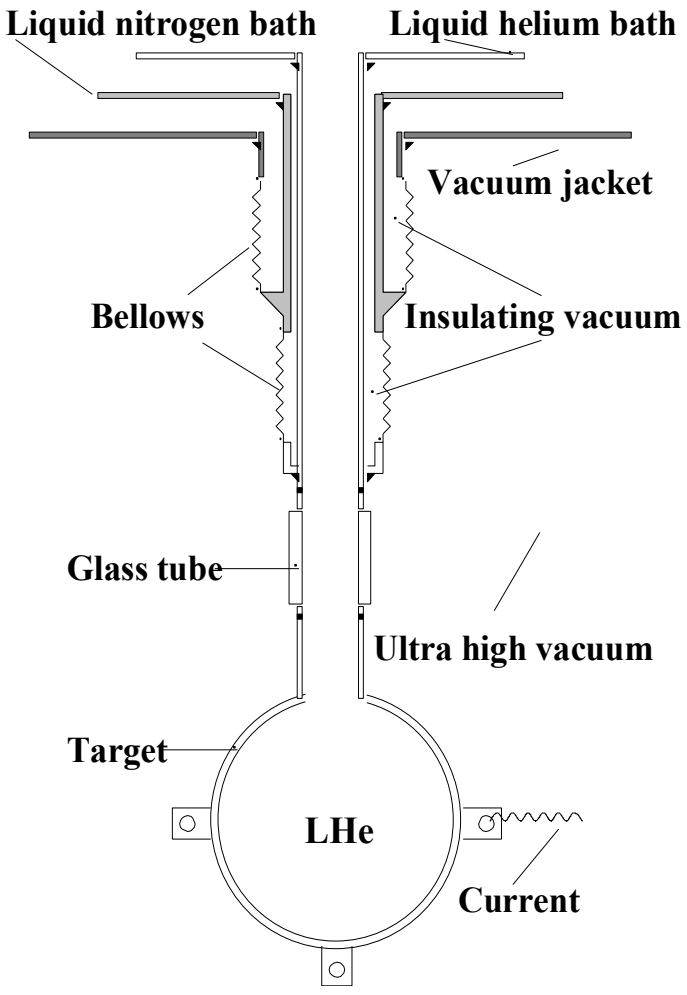

*Fig. 3.Schematic view of the target.*

An experimental arrangement for SEE measurements is shown on Fig.4 and it consists of an electron gun able to accelerate electrons until 3keV, a collector of electrons (cage), the revolving sample “dummy”, the target, beam driving and current measuring electronic equipment. The cage was biased at +45V relatively ground in order to prevent escape of secondary electrons from its surface.

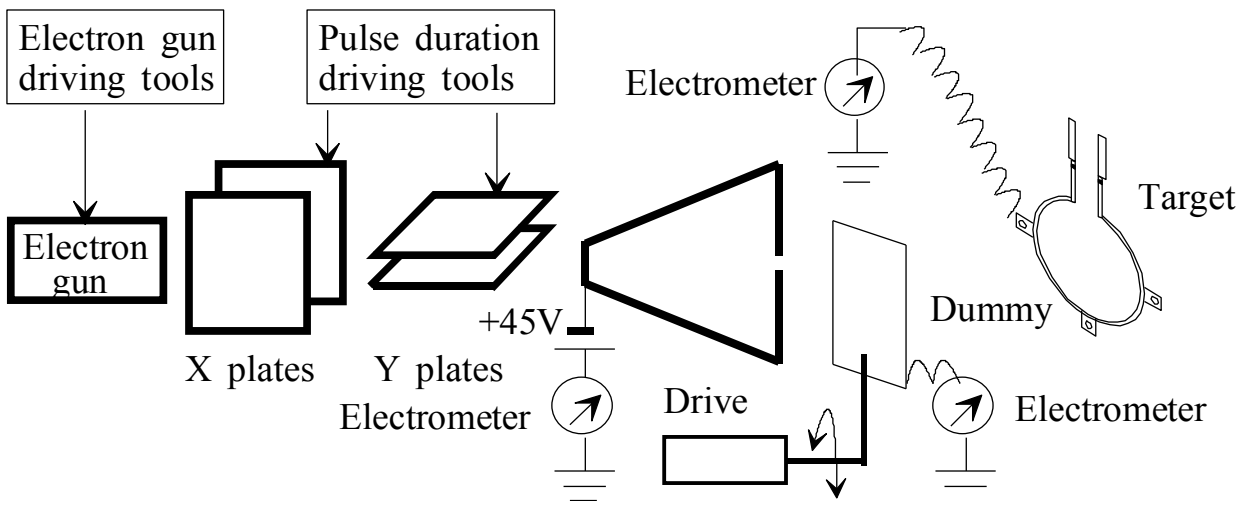

*Fig. 4.Diagram of the experimental arrangement.*

If $I_c$ is the current measured on the cage then the value of the secondary electron emission coefficient (SEEC) for the sample one can calculate as

$$\delta = \frac{I_c}{I_c + I_s} \qquad (1)$$

where

$$I_c + I_s \equiv I_b \qquad (2)$$

represents the beam current. The equation (1) can be rewritten in terms of electron doses acquired by the cage and the sample as

$$\delta = \frac{\int_0^\tau I_c(\tau)d\tau}{\int_0^\tau I_c(\tau)d\tau + \int_0^\tau I_s(\tau)d\tau} \qquad (3)$$

A 12 bits board MIO-16-L/9 from National Instruments with 10 µsec analog-to-digit conversion time was used to measure the currents and to drive the beam. Beam pulsing was carried out by means of an analog switch introduced in a deflection plate circuit. The switch toggles its output between “Beam-off” and “Beam on” voltages depending on level of a digital pulse coming from a counter of the board (Fig. 5). As effect the duration of the digital pulse defines the duration of the beam pulse. At the same time the digital pulse also triggers acquisition for the channels of current.

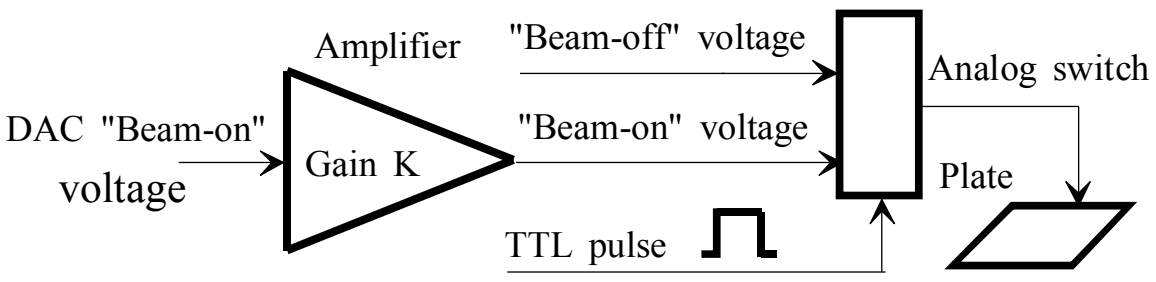

*Fig. 5. The beam pulse driving electronic scheme*

Current amplifiers Keithley 427 were used as electrometers. The filter time constant of the electrometers was chosen to be 0.1msec that allowed to register the maximal attained currents for the beam pulse duration down to 250µsec.

## Problems faced during measurements

It will not be useless to mention about problems faced during measurements since some of them are very learning. Already first attempts of measurements have shown that the SEEC for rare gases is strikingly high.

These attempts were carried out on the set-up not having a "dummy" sample installed and as consequence the beam current was defined via the equation (3) i.e. via the measurements of the cage and the sample currents. It was immediately seen that accuracy of the measurements is not good and has to be significantly improved. An explanation is in the fact that for the high values of the SEEC both the sample and the cage currents become of order of $\delta \cdot I_b$ having at the same time opposite signs. It results that an error in determination of the beam current could reach values as high as $2 \cdot \delta \cdot \theta$ , where $\theta$ is an error of measurements for the sample and the cage currents. Thus, if $\theta$ is of order of 1% then the error in determination of the beam current can reach 60% for a surface having SEEC=30. The problem was overcome after having introduced in the set-up a copper "dummy" sample serving for determination of the beam current. This has considerably improved the accuracy of the measurements since the SEEC for copper is much lower and the beam current can be measured via the DC method increasing signal-to-noise ratio during the measurements. Following this, the equation (3) was changed to

$$\delta = \frac{\int_0^\tau I_c(\tau)d\tau}{I_b \Delta\tau} \qquad (4)$$

where $\Delta\tau$ is the beam pulse duration and $I_b$ is the beam current measured on the copper sample accordingly to the equation (2).

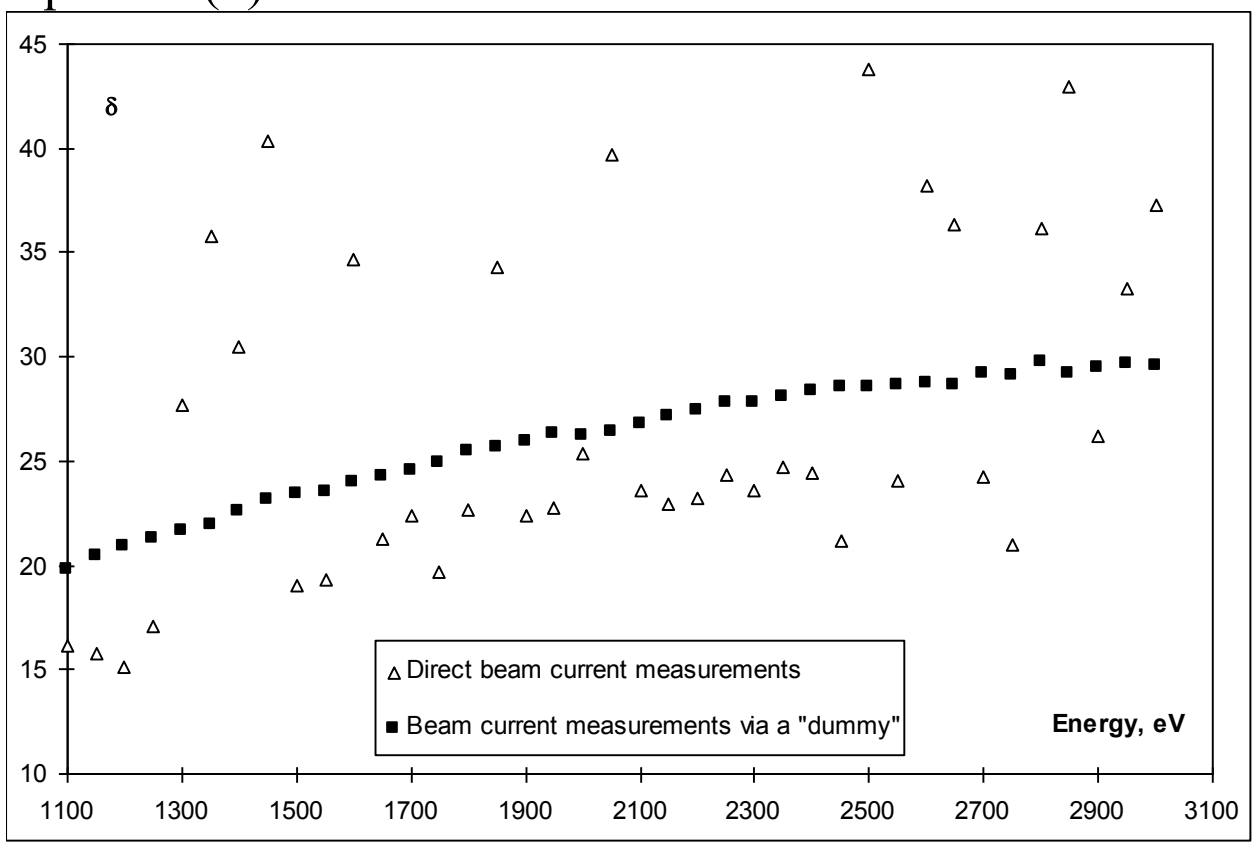

*Fig. 6. Comparison of the two different methods of the SEEC measurements.*
*- s- direct beam current measurements according to the equation (3)*
*- n - beam current measurements via the "dummy" according to the equation (4).*

The comparison of these two methods of the SEEC measurements is illustrated on Fig. 6. The results have been obtained for 1500 monolayers of argon with the beam pulse duration of 500μsec. One has to note that the shown data are only for comparison of the methods of the measurements and should not be considered as the correct ones.

The increase of the accuracy of the SEEC measurements has helped to understand an influence of the residual gases in the vacuum chamber on behavior of the SEEC. Since at the most beginning the penetration depth for the incident beam was not known it was thought that the dosing time could reach orders of a few hours. This period is great enough for deposition of few monolayers of a residual gas on the target which could considerably influence the correct SEEC measurements. The results of the undertaken research are shown on Fig. 7.

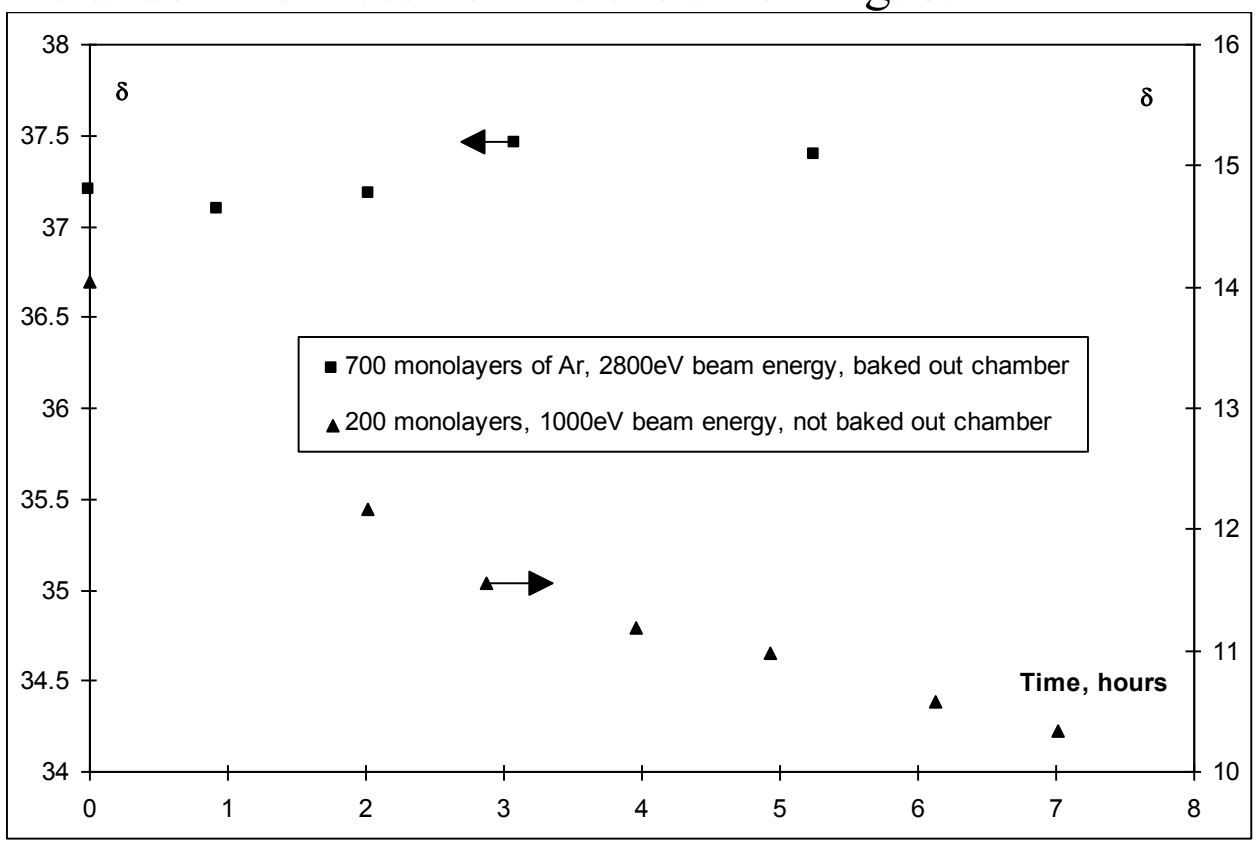

*Fig. 7. Influence of the initial pressure in the vacuum chamber on the SEEC measurements.*

First attempts have been carried out with initial pressure in the vacuum chamber of $1.0 \bullet 10^{-9}$torr coming to $6.0 \bullet 10^{-10}$torr after having filled up the cryostat with liquid helium. In order to avoid a possible influence of the surface charge only one 500μsec beam pulse per hour has been allowed with the beam current of $1.0 \bullet 10^{-9}$A. The results obtained for a modest coverage of 200 monolayers of argon for the beam with energy of 1000eV have shown an unacceptable influence of residual gases in the vacuum chamber on the SEEC measurements. As a consequence the vacuum chamber has been then baked out at $300^0$C during 24 hours what has resulted in the initial pressure decrease down to $1.0 \bullet 10^{-10}$torr and coming to the $5.0 \bullet 10^{-11}$torr after filling up the cryostat with liquid helium. The results obtained for the coverage of 700 monolayers of argon for the beam with the energy of 2800eV have already been considered as satisfactory ones and all following SEEC measurements have been therefore carried out with the initial pressure in the vacuum chamber of $1.0 \bullet 10^{-10}$torr.

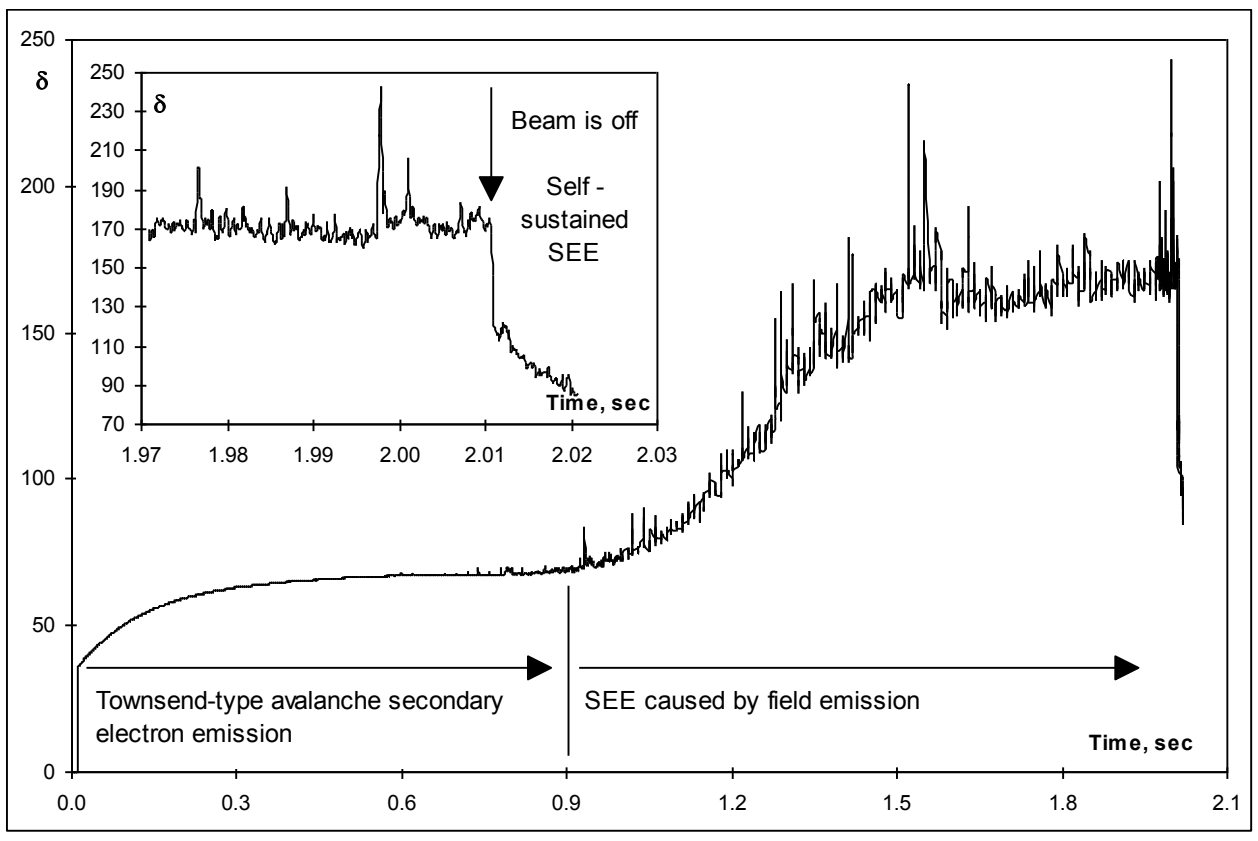

*Fig.8. Secondary electron yield as function of the bombardment time obtained on 700 monolayers of argon with the beam current being of $2.9 \bullet 10^{-9}$A.*

Next step to be understood was the well known charge problem. It was supposed that because of a relatively small thickness of a deposited gas, the internal field across the film will be rather first dominating mechanism limiting achievement of the correct results than the field between the surface of the film and the collector of electrons. The behavior of the SEEC for every gas has been verified via irradiation of fresh deposited films with 3keV energy electron beam during 2 seconds with simultaneous measurements of the sample and cage currents. The sampling rate has made up 100μsec, the beam current was measured via the "dummy". The results of these measurements for argon, neon, krypton and xenon are shown on Figures 8-11.

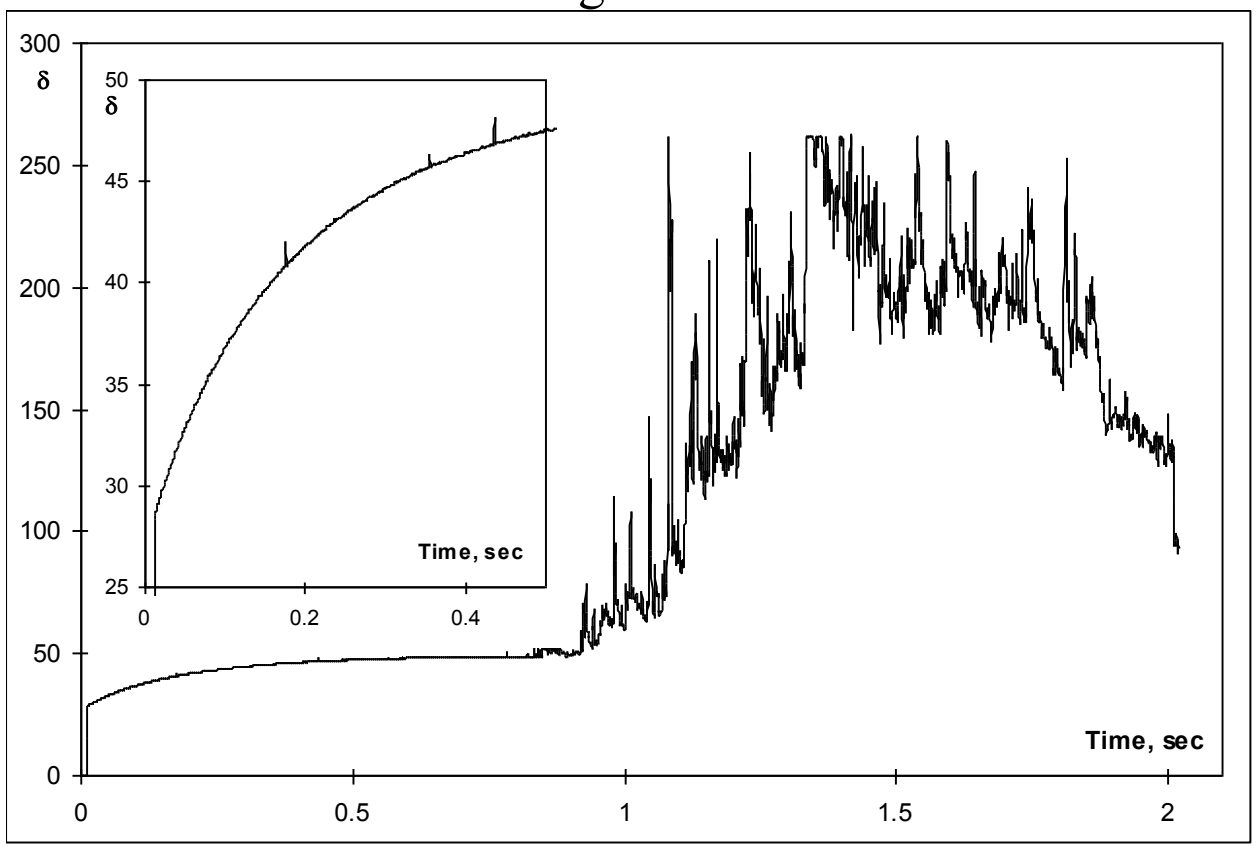

*Fig.9. Secondary electron yield as function of the bombardment time obtained on 1000 monolayers of neon with the beam current being of 3.8•10$^{-9}$A.*

.

In general, under these testing conditions all gases have exhibited about the same behavior. The surface of the film acquires a positive charge just after beginning of the bombardment since the SEEC>>1. If the film of a condensed gas has a porous structure what is as it is seen obviously the case for all tested gases, then the appearing field across the film accelerates secondary electrons during their travel to the surface mainly via pores of the film to an energy sufficient to liberate new electrons.

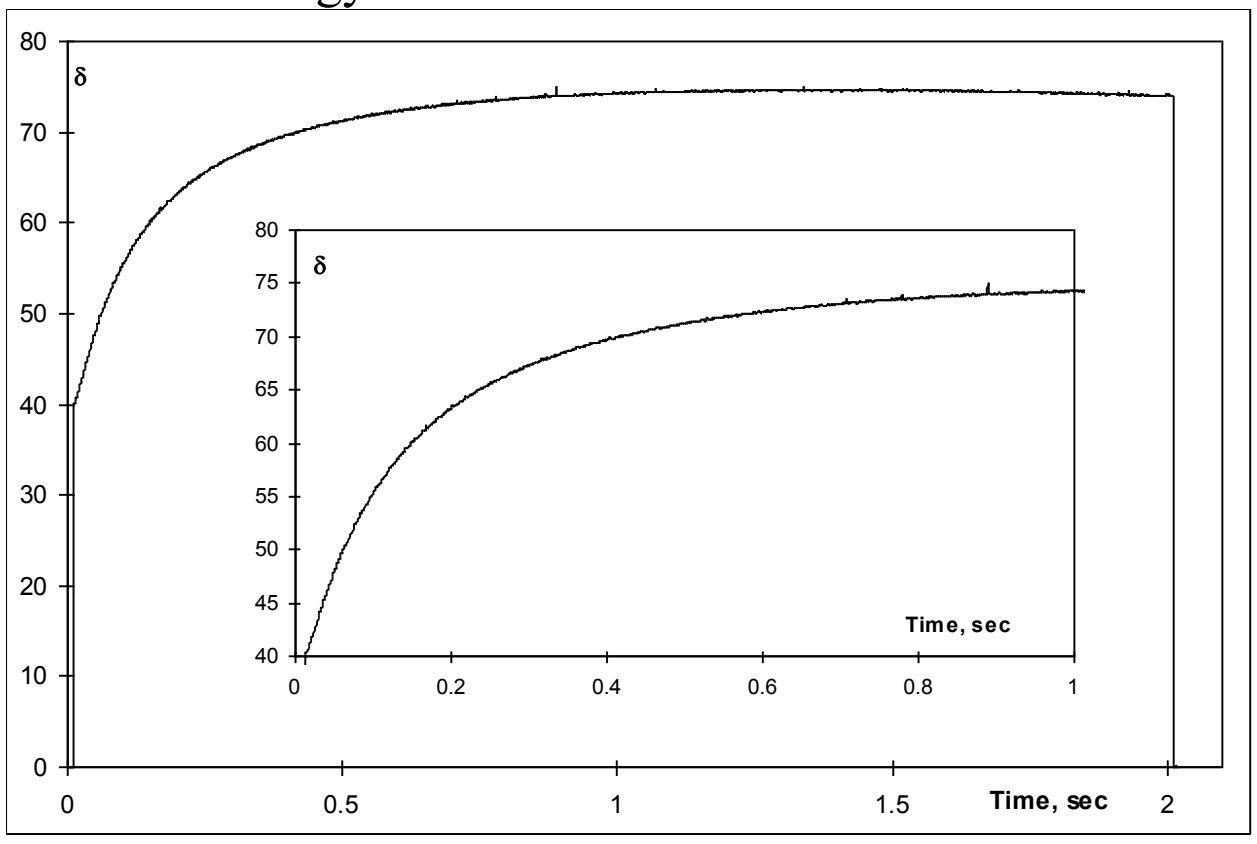

*Fig.10. Secondary electron yield as function of the bombardment time obtained on 500 monolayers of krypton with the beam current being of 2.4•10$^{-9}$A.*

These electrons create new ones by ionization and a Townsend-type avalanche or stimulated electron emission results[2]. A stable value of the electron yield is reached when the surface potential approaches to the collector potential. The stimulated secondary electron emission and its approaching to a stable value is observed for all tested gases after beginning of the bombardment.

With the surface potential going up due to continuous electron bombardment and removal of secondary electrons, at a certain moment the field gradient across the film can become so high that results appearance of field emission. This phenomenon known as Malter effect[3,4] or field emission stimulated secondary electron emission was observed during these experiments for argon and neon. It can be seen that the films still emitted the electrons for a certain time even after the stopping the bombardment.

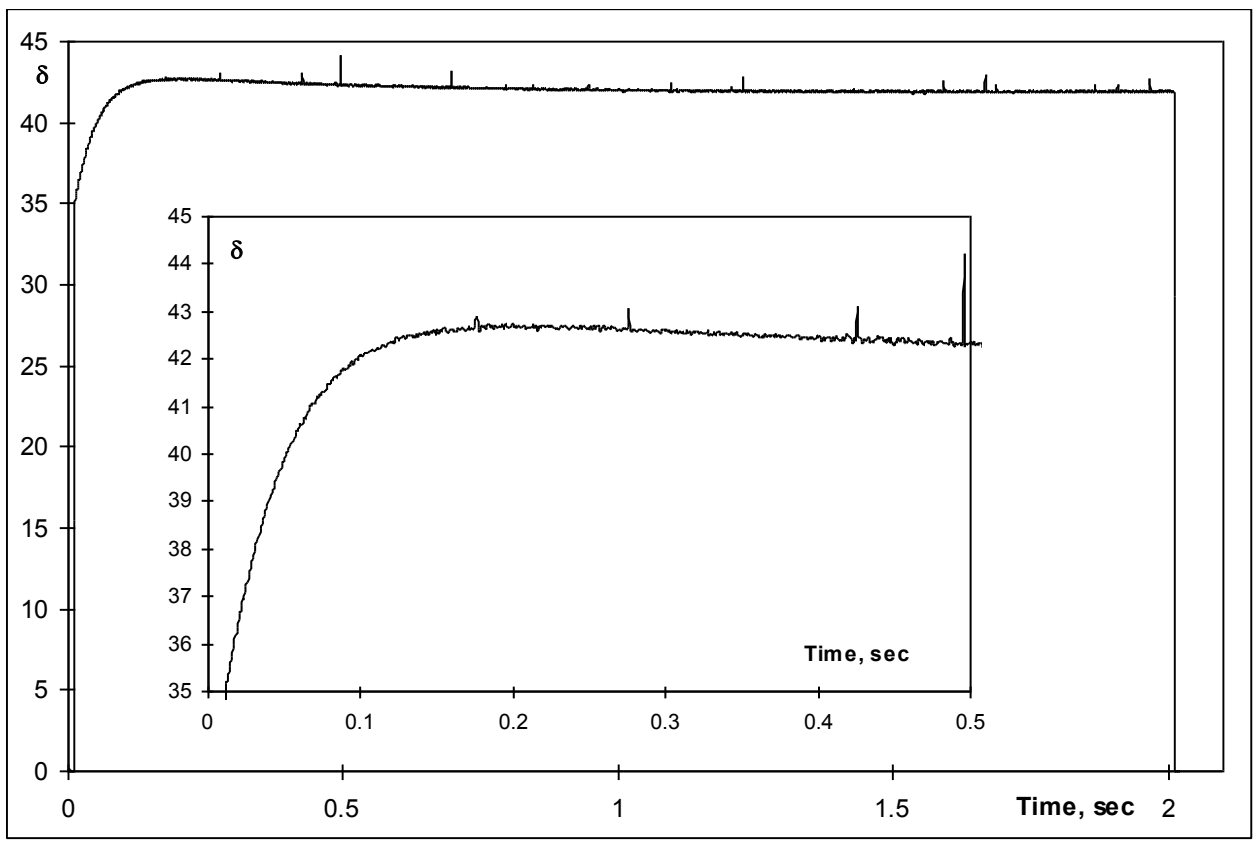

*Fig.11. Secondary electron yield as function of the bombardment time obtained on 400 monolayers of xenon with the beam current being of 3.4•10$^{-9}$A.*

These experiments have shown an importance of a correct choice of a beam dose. As a result the beam current being the same for the all SEEC measurements was chosen to be $I_b$ =9.0•10$^{-10}$A. An exception has been allowed while measuring the SEEC for low coverages where the beam current made up about 5.0•10$^{-9}$A. A few attempts have been carried out for determination of the beam pulse duration. The most presented further SEEC data have been obtained with the beam pulse duration of 500μsec. The highest values of the SEEC (>40) have been measured with the lower pulse duration till 250μsec.

Each time after having bombarded the target with electron beam it is necessary to renew the adsorbate. Since a goal of this research was to obtain the dependence of the SEEC from the incident energies ranging from 100eV till 3keV and the coverages for 4 rare gases, renewing the film after each beam impact would put important problems related to high consumption of liquid helium and unacceptable time to finish this research. A solution for speeding up the measurements has been found in doing the SEEC measurements in direction from high energies to the low ones. An illustration for this can be seen on Fig. 13 where each time renewed 4000 monolayers of neon have been measured in different directions: from 100eV to 3keV, from 3kev to 100eV and from 2keV to 1keV. One can see a significant difference in results for 3keV obtained on a fresh (3keV→100eV) and charged (100eV→3keV) layers. Conversely, no difference was found in results for 2keV comparing the fresh and charged layers for direction from high to low energies. Further attempts with different initial energies for different gases have confirmed the correctness of this solution. This fact that low energy incident electrons do not "see" consequences of the previous beam impact with electrons of higher energy has been a surprising and pleasant gift from nature helping in speeding up the SEEC measurements for rare gases. All presented further the SEEC results have been obtained for the direction from 3keV to 100eV. It means that for once deposited layer of a certain thickness, the SEEC dependence from incident energy was measured for whole range of energies starting from 3keV to 100eV without renewing the film.

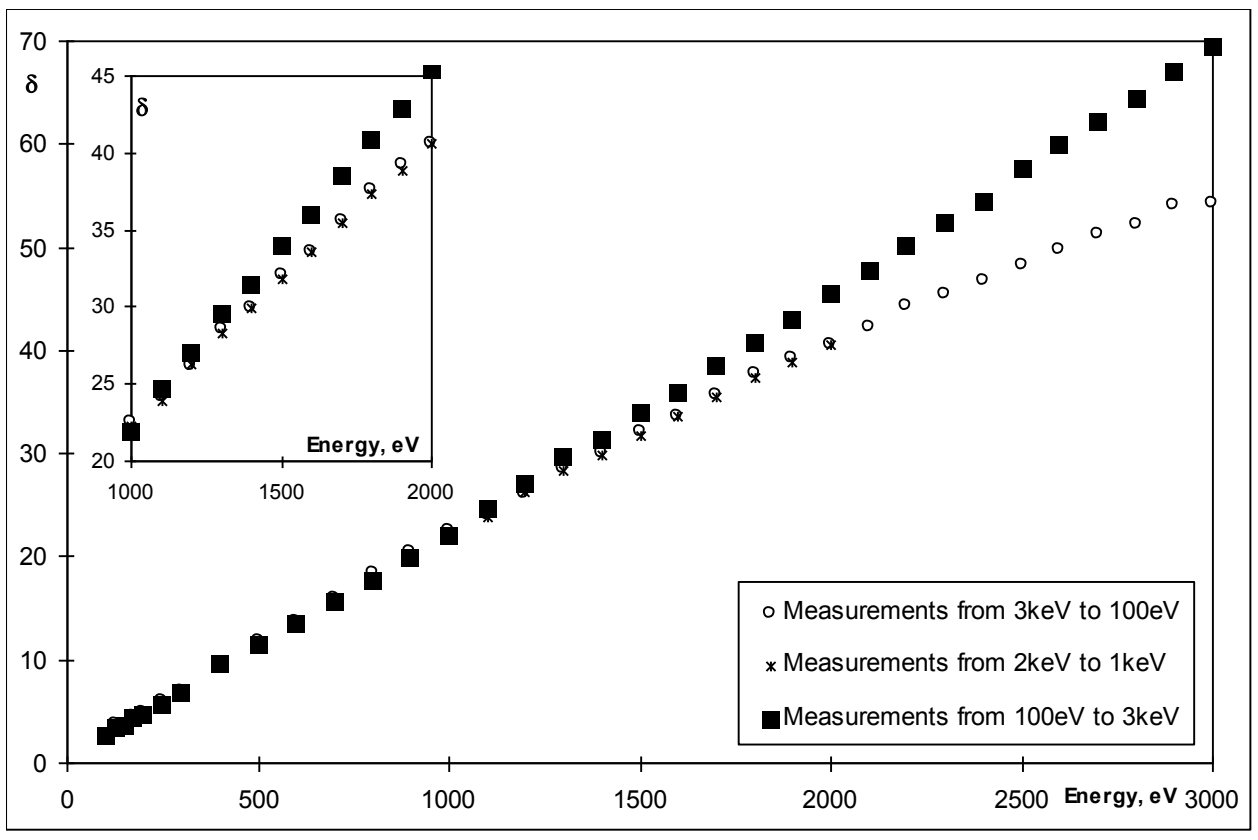

*Fig.13. Comparison of the SEEC results obtained on 4000 monolayers of neon depending on direction of measurements..*

A charge acquired by a surface during the beam impact can be released either via flooding the surface with low energy electrons from a neighboring filament or via evaporating of the film of a condensed gas with the help of a heater and deposition of a fresh layer. The present set-up was not equipped with such tools that is why another solution has been applied to evaporate the film (Fig. 12). Warm helium was blown via a thin wall tube inserted in the cryostat and going till the bottom of the target. This tube was surrounded with another tube external diameter of which was reduced for a part going to the neck of the target. This transient part between two diameters, having a form of a cone served as a plug isolating the liquid helium in the main cryostat from the target during evaporation. The evaporated helium goes upwards in the space between the tubes in recuperation system. All assembly was used only for the period needed to evaporate a film, liberated gas was pumped down with the turbomolecular pump. Complete removal of the film was controlled via the SEEC measurements comparing them with the ones for clean target.

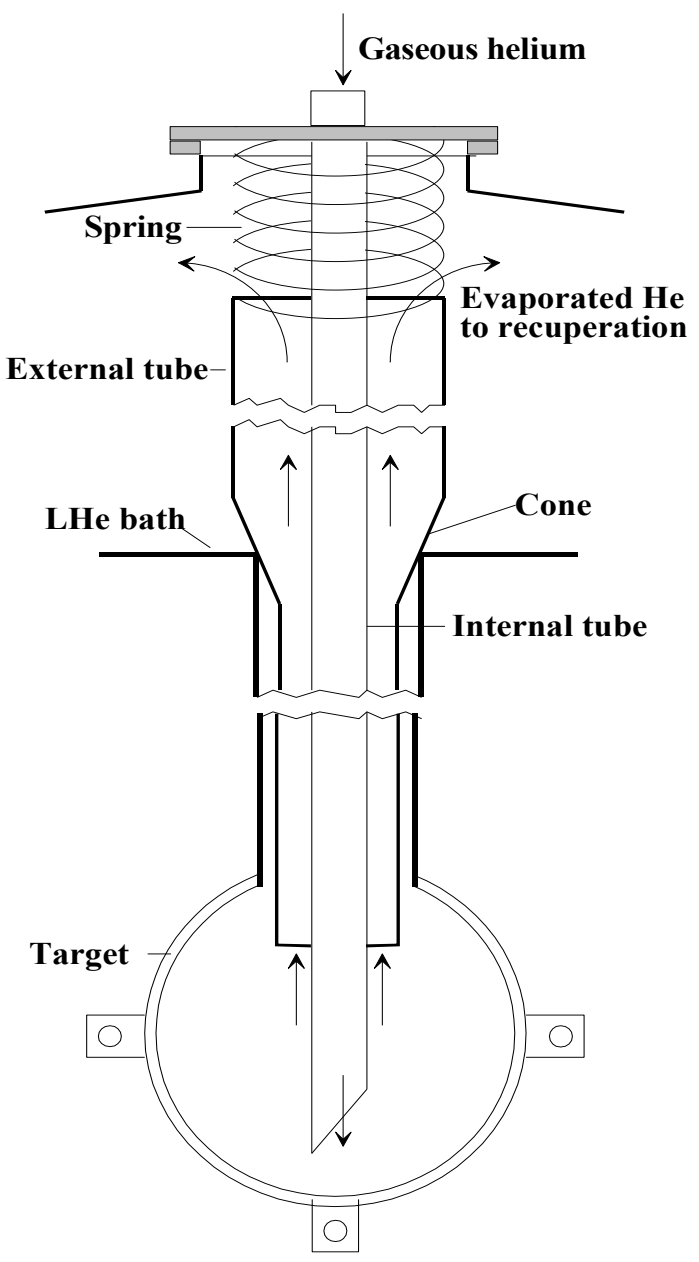

*Fig. 12. Arrangement for the film removal.*

## Procedure of measurements

The vacuum chamber together with the injection line have been once baked out at $300^0$C during 24 hours. The attained pressure was about $1.0•10^{-10}$torr as it was mentioned above. Each change of bottle of the gas to be investigated was followed with baking out of whole system at $150^0$C. Before doing first injection the injection line was rinsed 4-5 times.

The pressures before diaphragm and in the vacuum chamber during injection were kept about the same for all gases and made up correspondingly about $9.5•10^{-5}$torr and $3.0•10^{-6}$torr of nitrogen equivalent. The injection was carried out with the main turbomolecular pump valved off, the cold surface of the target estimated as being of $25cm^2$ was the only pumping tool for the injected gas. The "dummy" sample during injection was put in the most far position from the cold target in order to allow a homogeneous deposition of a film on whole cold surface. Once the deposition of a required film thickness was completed the "dummy" sample was placed in the position just in front of the target and adjustment of the gun for whole range of energies was carried out with simultaneous measurements of the beam current. Then, again the "dummy" sample was put in the most far position from the target and the SEEC measurements in direction from 3keV to 100eV were carried out on the deposited film. The gun adjustment and the SEEC measurements procedure take about 15 minutes. As soon as all measurements are taken the film evaporation procedure is executed with simultaneous pumping of the liberated gas with the turbomolecular pump. After having finished the evaporation, a film of another thickness can be deposited on the target with following repetition of all above mentioned steps.

**Table 1. Summary of some data for investigated gases.**

| | Neon | Argon | Krypton | Xenon |
|---|---|---|---|---|
| Purity index | 40 | 57 | 48 | 45 |
| Purity, % | 99.990 | 99.9997 | 99.998 | 99.995 |
| Molecules/ monolayer | $1.0•10^{15}$ | $6.9•10^{14}$ | $6.14•10^{14}$ | $5.33•10^{14}$ |
| Atomic weight | 20 | 40 | 84 | 132 |
| Sensibility of vacuum gauge | 4.1 | 0.8 | 0.5 | 0.4 |
| Dosing rate, monolayers/sec | 4.8 | 0.99 | 0.48 | 0.35 |

The Table 1 summarizes some data for the investigated gases. The expression "sensibility of vacuum gauge" means a sensibility of the gauge for a given gas relatively nitrogen.

## Results of the SEEC measurements for rare gases

As it was already stated , the goal of this research was to obtain 3-D dependence of the SEEC from incident energy and coverage for neon, argon, krypton and xenon. The range of incident energies depended of the electron gun possibilities and was the same for all gases. In total, 34 values of the incident energy have been used. They ranged from 3000eV down to 300eV with the step of 100eV and then 250, 200, 175, 150, 125 and 100eV. The maximum value of applied coverage depended of the fact when influence of the copper substrate on the SEEC became negligible. That is, the SEEC measurements for this coverage could be considered as the ones belonging only to the explored gas.
Since it was not possible to show all obtained results, an extraction of results for every gas giving an idea about the SEEC dependence from incident energy and coverage will be presented. An array of the results for each gas includes the SEEC dependence from incident energy for the clean cold target, corresponding to the coverage in 0 monolayers.

*Neon.*

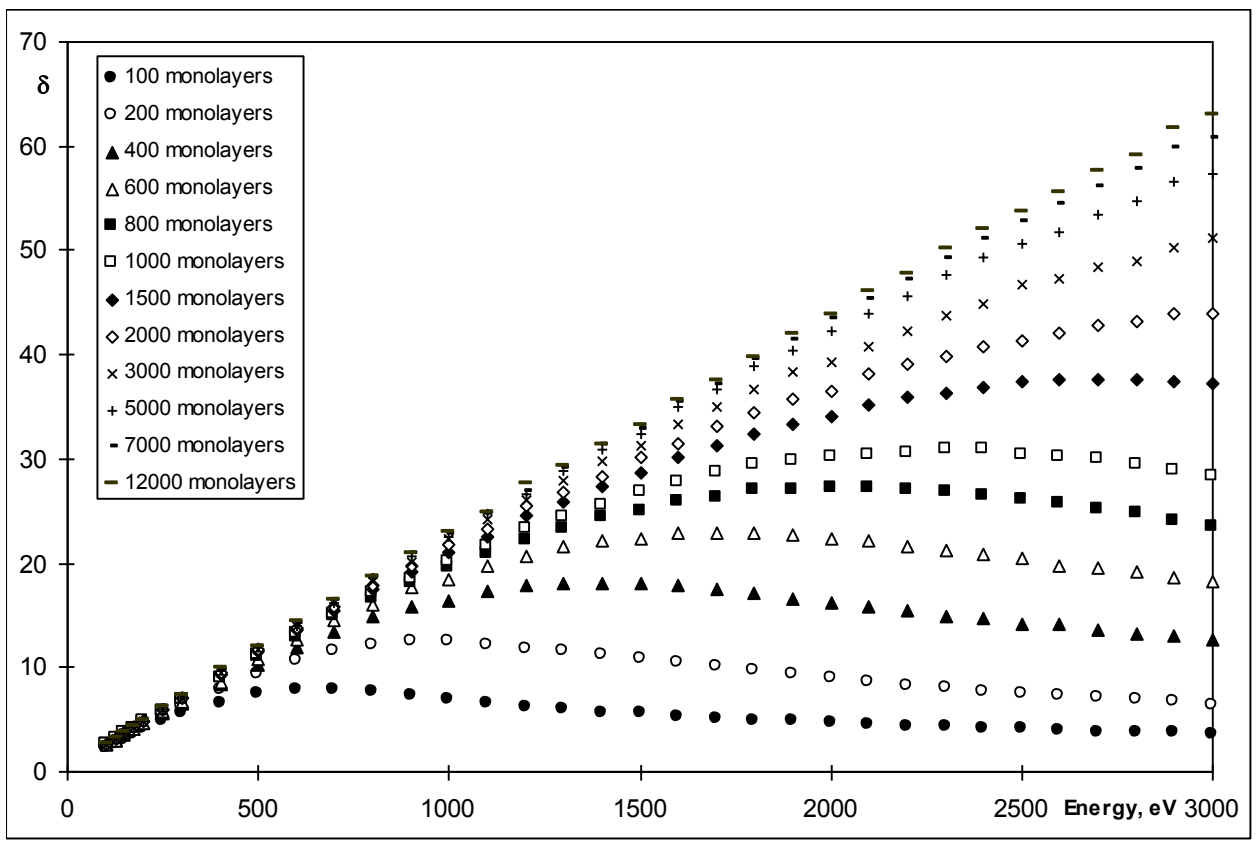

*Fig. 14. "Secondary electron yield - incident energy" plane for neon.*

Neon was explored for the following 27 different coverages: 28, 40, 60, 80, 100, 150, 200, 300, 400, 500, 600, 700, 800, 900, 1000, 1200, 1500, 1700, 2000, 2500, 3000, 4000, 5000, 6000, 7000, 9000 and 12000

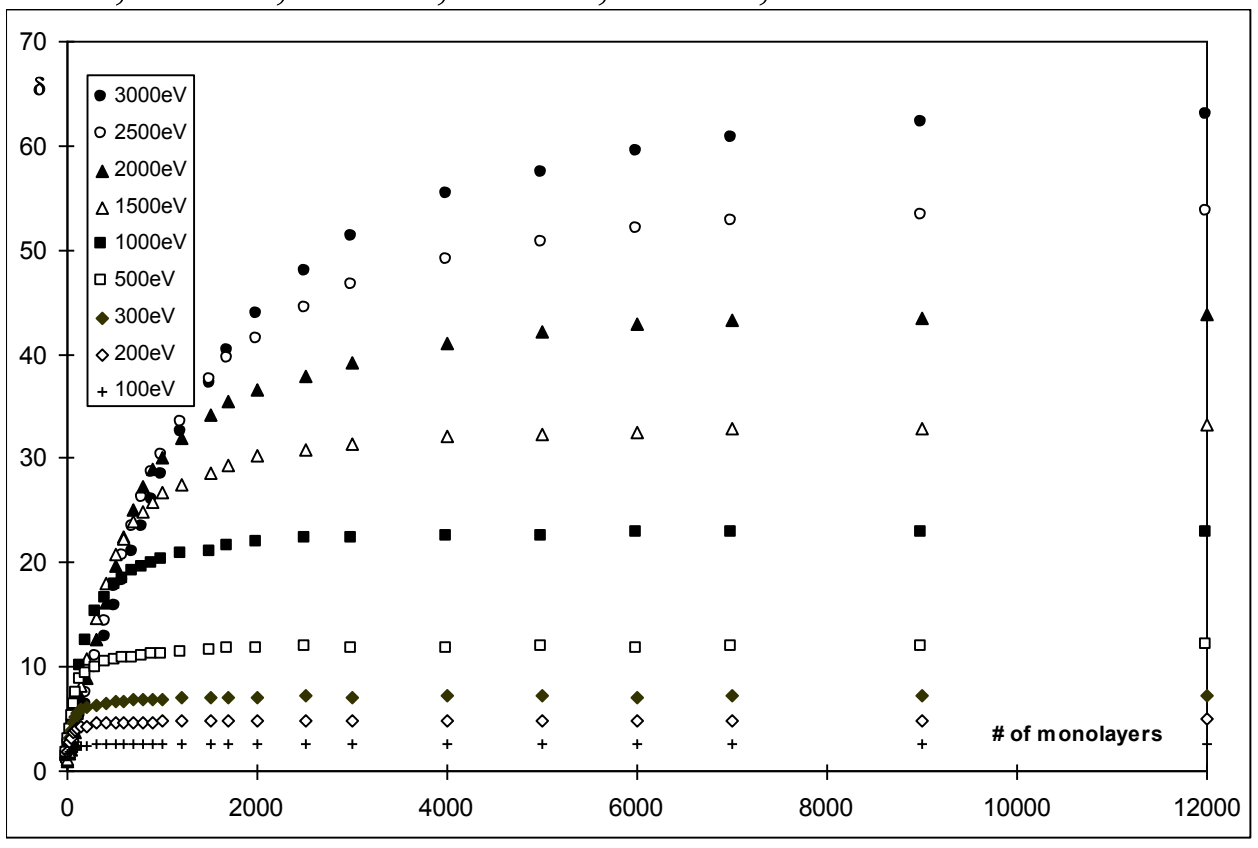

*Fig. 15. "Secondary electron yield - coverage" plane for neon.*

monolayers. The obtained results are shown on the Figures 14, 15. The maximal value of the SEEC was found about 62 for the incident energy of 3keV. Since the dependence of the SEEC as a function of the coverage seems to be of an exponential character, the penetration depth for a fixed energy can be taken as a value corresponding to a coverage where the SEEC reaches 63% of its equilibrium value. Thus, for neon the penetration depth for 3keV can be estimated as 1700 monolayers. The SEEC dependence from incident energy obtained for 12000 monolayers looks like quite straight line showing that the escape depth of the secondary electrons for neon is still far from this coverage.

*Argon.*

Argon was explored for the following 23 coverages: 20, 40, 60, 80, 100, 150, 200, 250, 300, 400, 500, 600, 700,

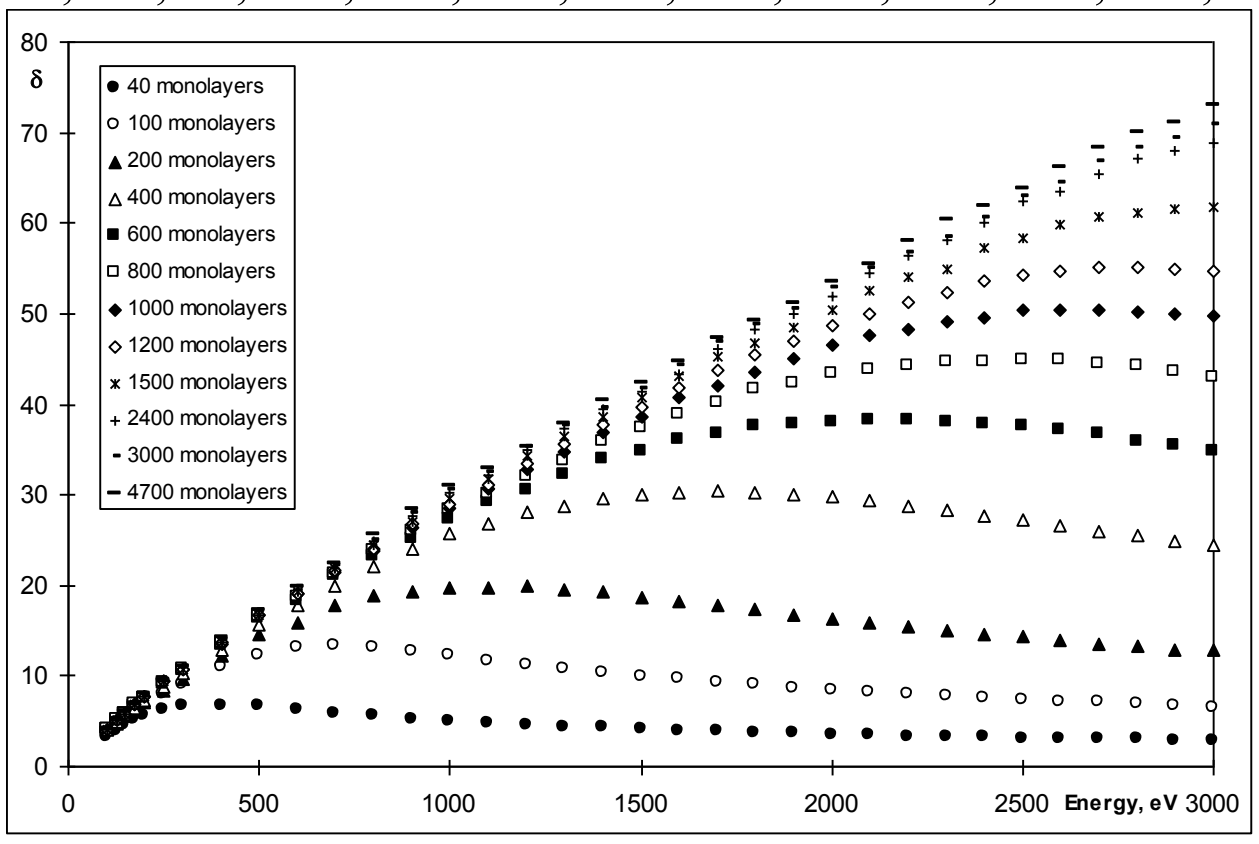

*Fig. 16. "Secondary electron yield - incident energy" plane for argon.*

800, 900, 1000, 1200, 1500, 1800, 2400, 3000, 3600 and 4700 monolayers. The obtained results are shown on the Figures 16, 17. The maximal value of the SEEC was found as 73 for the incident energy of 3keV and it is presently the greatest value for the investigated gases. The penetration depth for this energy could be estimated as of 890 monolayers. Again as for neon, the SEEC dependence from incident energy obtained for the last coverage in 4700 monolayers for argon looks like a straight line.

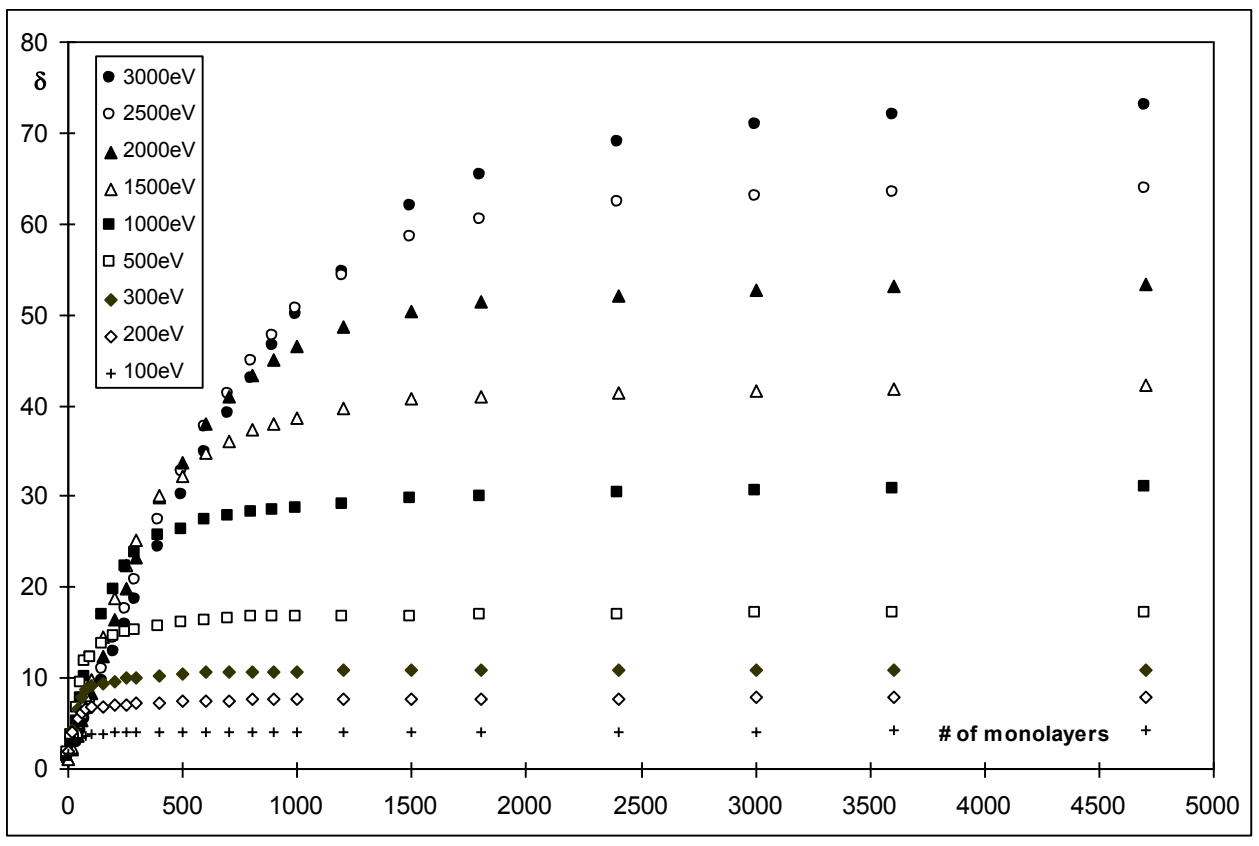

*Fig. 17. "Secondary electron yield - coverage" plane for argon.*

*Krypton.*

Krypton was explored for the following 20 coverages: 10, 20, 40, 60, 80, 100, 150, 200, 250, 300, 400, 500, 600, 700, 800, 1000, 1200, 1500, 2000 and 2500 monolayers. The obtained results are shown on the

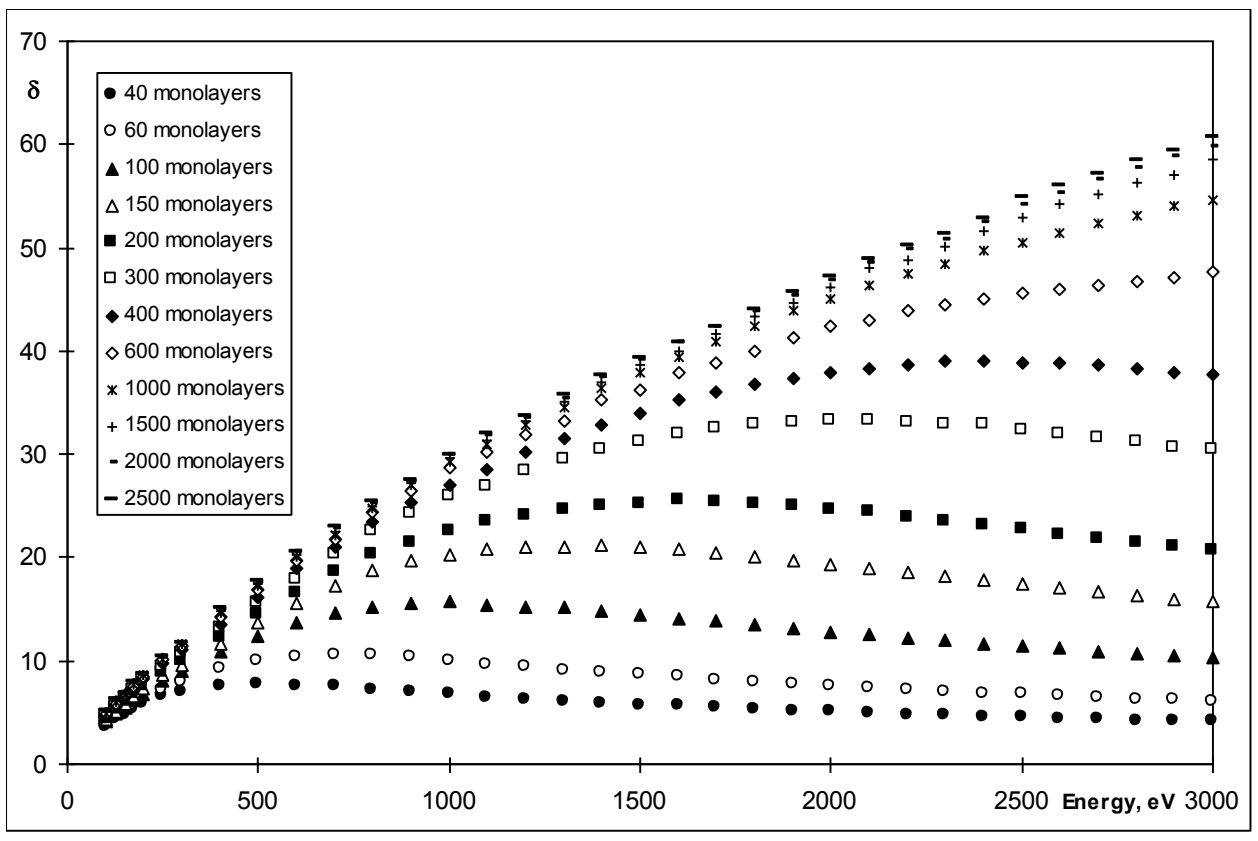

*Fig. 18. "Secondary electron yield - incident energy" plane for krypton.*

Figures 18, 19. The maximal value of the SEEC was found as 60.5 for the incident energy of 3keV, the penetration depth corresponding to this energy is about of 420 monolayers. One can already notice a deviation of the SEEC dependence from incident energy from straight line for the thickest used coverage of 2500 monolayers that could be attributed to a decrease of inelastic cross section for the higher energy incident electrons and to reducing the number of the escaping secondary electrons due to greater penetration depth.

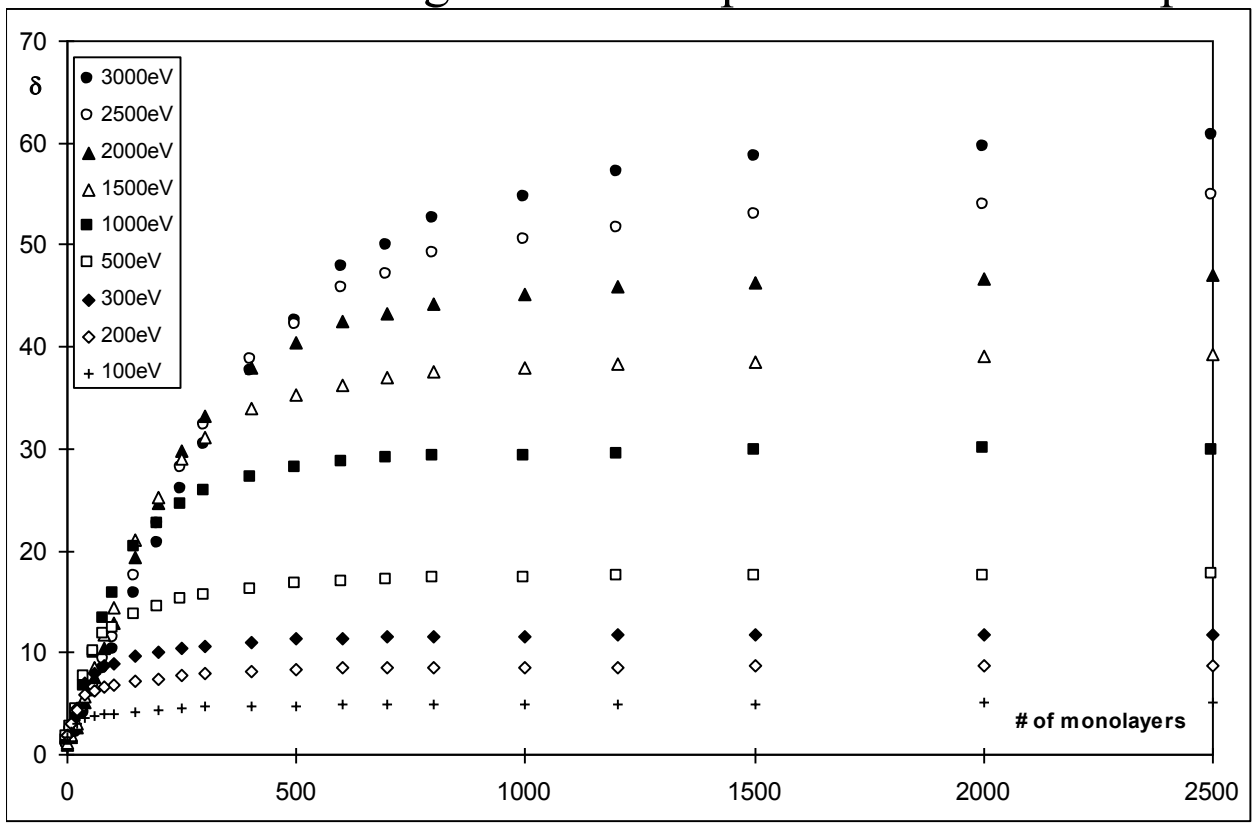

*Fig. 19. "Secondary electron yield - coverage" plane for krypton.*

<u>Xenon</u>

And the last gas, xenon was explored for the following 16 coverages: 10, 20, 40, 60, 80, 100, 150, 200, 300, 400, 500, 600, 700, 800, 1000 and 1400 monolayers. The results are presented on the Figures 20, 21. Again the maximal found value of the SEEC for xenon corresponded to the incident beam energy of 3keV and made up about 52. The penetration depth for this energy is estimated to be about 280 monolayers. Comparing to other gases the SEEC dependence from investigated range of incident energies for xenon exhibits greater deviation from straight line meaning more strong influence of inelastic cross section and probability for secondary electrons to escape from greater penetration depth.

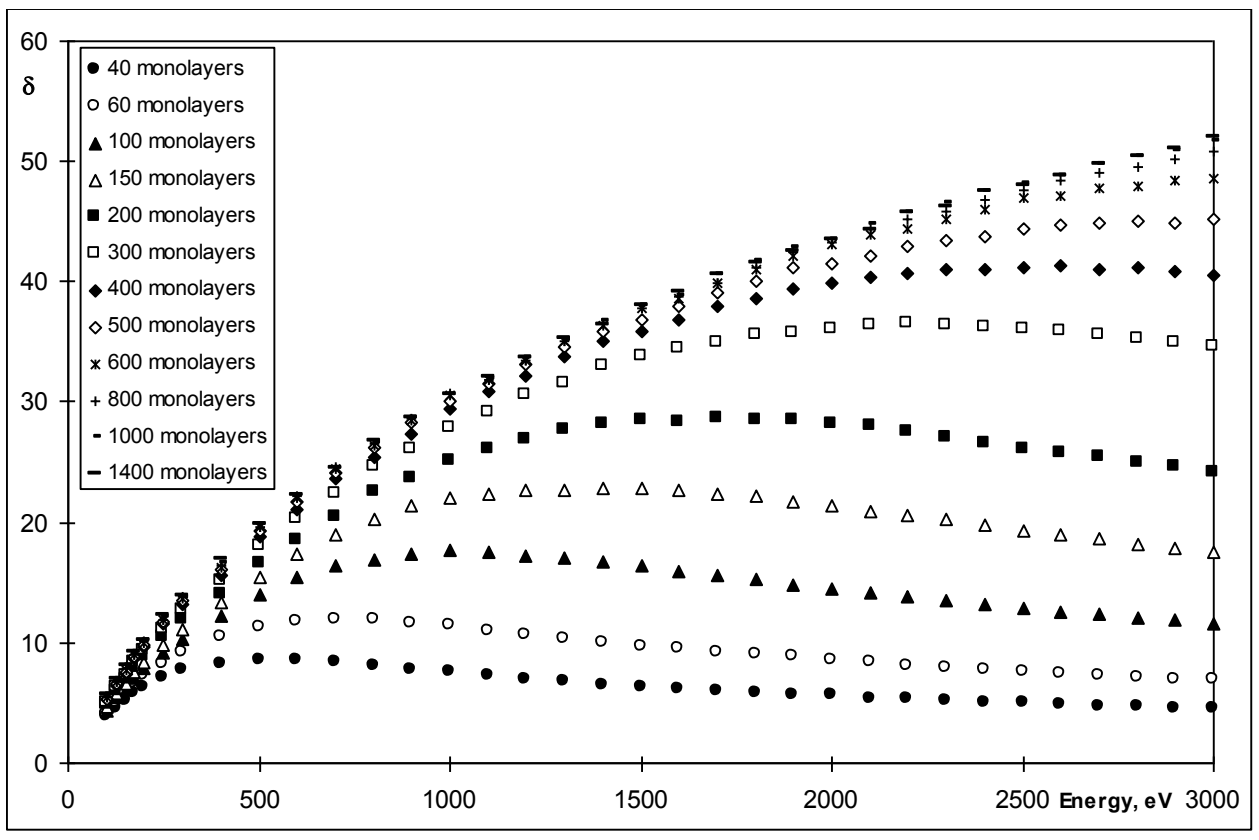

*Fig. 20. "Secondary electron yield - incident energy" plane for xenon.*

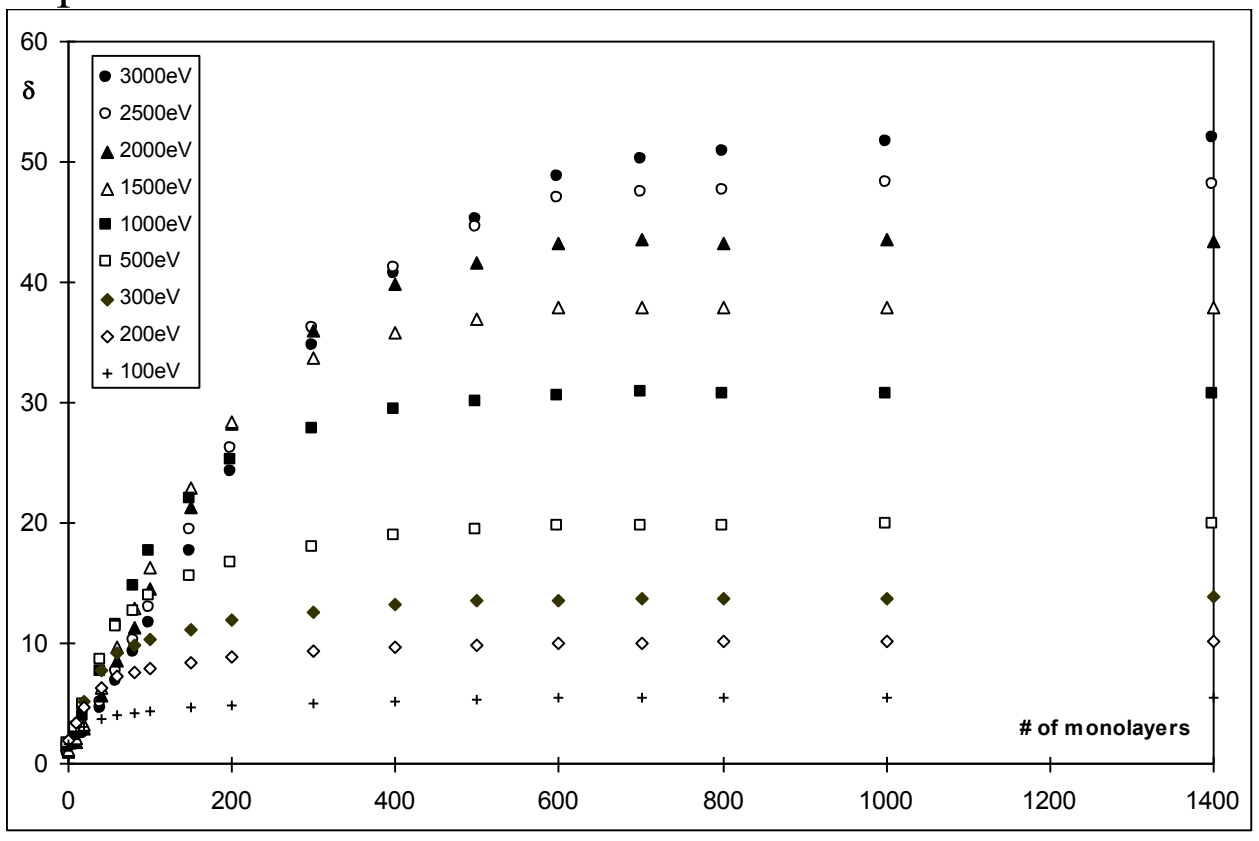

*Fig. 21. "Secondary electron yield - coverage" plane for xenon.*

Figure 22 summarizes the SEEC data for the investigated gases. Here, the data for every gas have been taken for its highest used coverage and then, could be considered as the SEEC data belonging only to the explored gases without taking into account influence of the substrate.

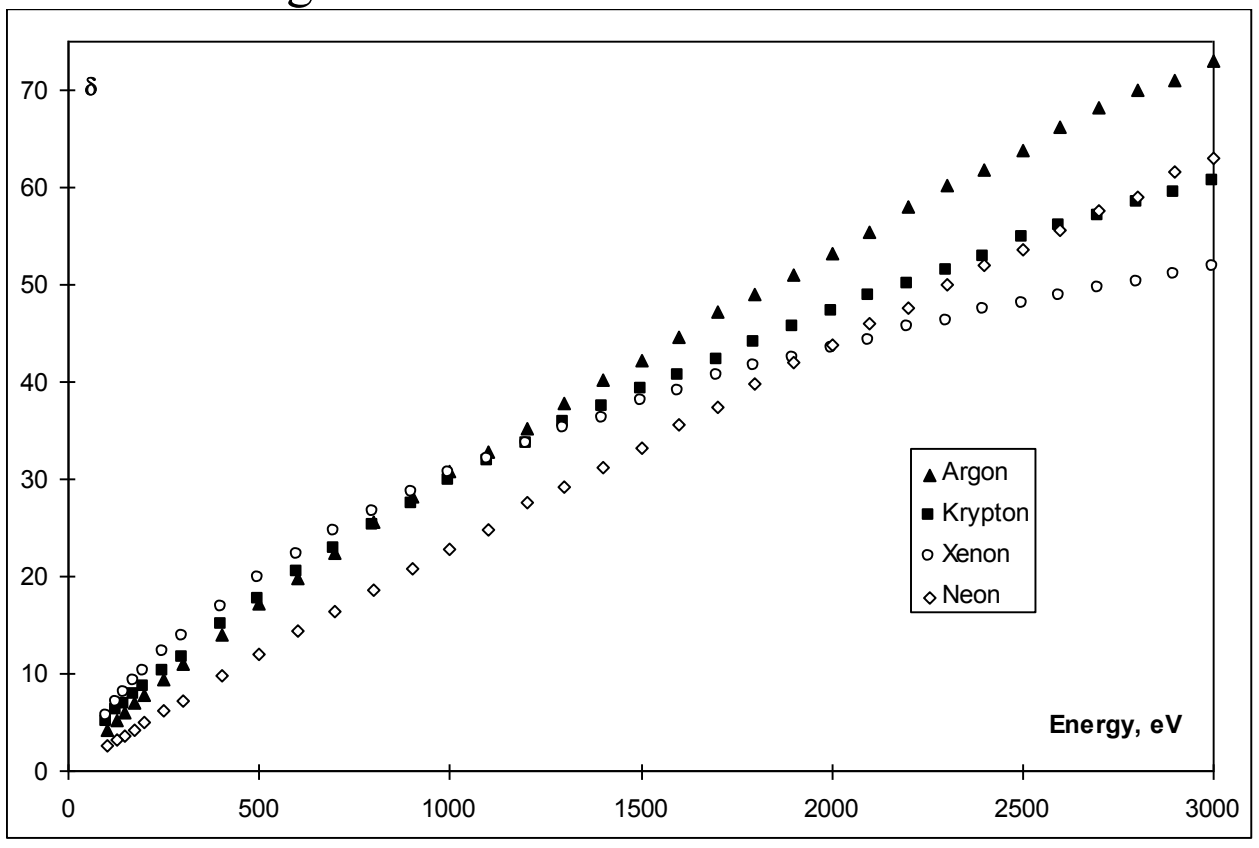

*Fig. 22. The SEEC as function of the primary energy for rare gases.*

## Measurements of the first cross-over.

Behavior of the SEEC as a function of the energy of the incident beam for all investigated gases showed that the values of energies for the first cross-over should lie in enough low energy region where the problems of the beam divergence start to play an important role for correct measurements of the secondary electron emission and require specially designed apparatuses for it[5]. But the measurements of the first cross-over can be significantly simplified if to control a value of the beam incident energy when the sample current falls to zero[6]. This method was applied for the measurements of the first cross-over for all investigated gases. In order to avoid the beam divergence while passing the beam through the gun, the sample and "dummy" were biased at -22.5V relatively ground.

**Table 2. Values of energies for the first cross-over.**

| | Neon | Argon | Krypton | Xenon |
|---|---|---|---|---|
| Energy, eV | 21 | 14 | 12.5 | 10.5 |

The procedure of the determination of the first cross-over was the following. After having measured the SEEC dependence from incident energy for a certain coverage, a few beam pulses at certain low energies were carried out in order to have an idea about range of energies where the first cross over could lie. With each renewal of the film to the thicker one the procedure was repeated with following narrowing of uncertainty in the energy range. And finally, when the uncertainty made up 0.5eV, one control beam pulse was carried out for a fresh and enough thick layer. One has to note that for a purpose of increasing of accuracy while doing the first cross-over measurements, the beam current was raised up to $1.5 \cdot 10^{-8}$A. The measured values of the first cross-over are summarized in the Table 2.

## Discussion

The SEEC results presented on the Fig. 22 show that the heavier gas the more the SEEC for the low energy region. The speed of the SEEC increase per monolayer at 100eV has roughly made up the following values: 0.005 for neon, 0.0596 for argon, 0.0688 for krypton and 0.0778 for xenon. However, the heavier gases due to stronger influence of the inelastic cross section decrease for high energy incident electrons will exhibit less maximal value SEEC occurring at a less incident energy in comparison with the lighter gases. Thus, one can expect that neon in its final SEEC dependence from incident energy will exhibit the highest maximal value of the SEY among rare gases.

Regarding to the SEEC results as a function of the coverage it is difficult to take decision about a value of penetration depth for a fixed beam energy because of an exponential character of the obtained curves . As it was already mentioned, a value playing a role of the dumping constant in an exponential fit $e^{-R/Rp}$ of these curves could be taken as a "penetration depth constant" $R_p$. Correctness of this can be evaluated with the help of an equation[7] describing a practical range of electrons in $Al_2O_3$:

$$R_p = 0.0115 \cdot E_0^{1.35} \quad (5)$$

where $E_0$ is an incident beam energy.

Calculations carried out for the 3keV incident energy give the values of the range for neon, argon, krypton and xenon correspondingly as 1200, 850, 456 and 315 monolayers. Comparing these results with the ones derived from the measured SEEC curves one can see that agreement is satisfactory.

It is interesting to mention that taking the range data from the measured SEEC results and plotting them as a function of the atomic number reciprocal $Z^{-1}$ one can observe that for high energies the plot looks as a quite straight line.

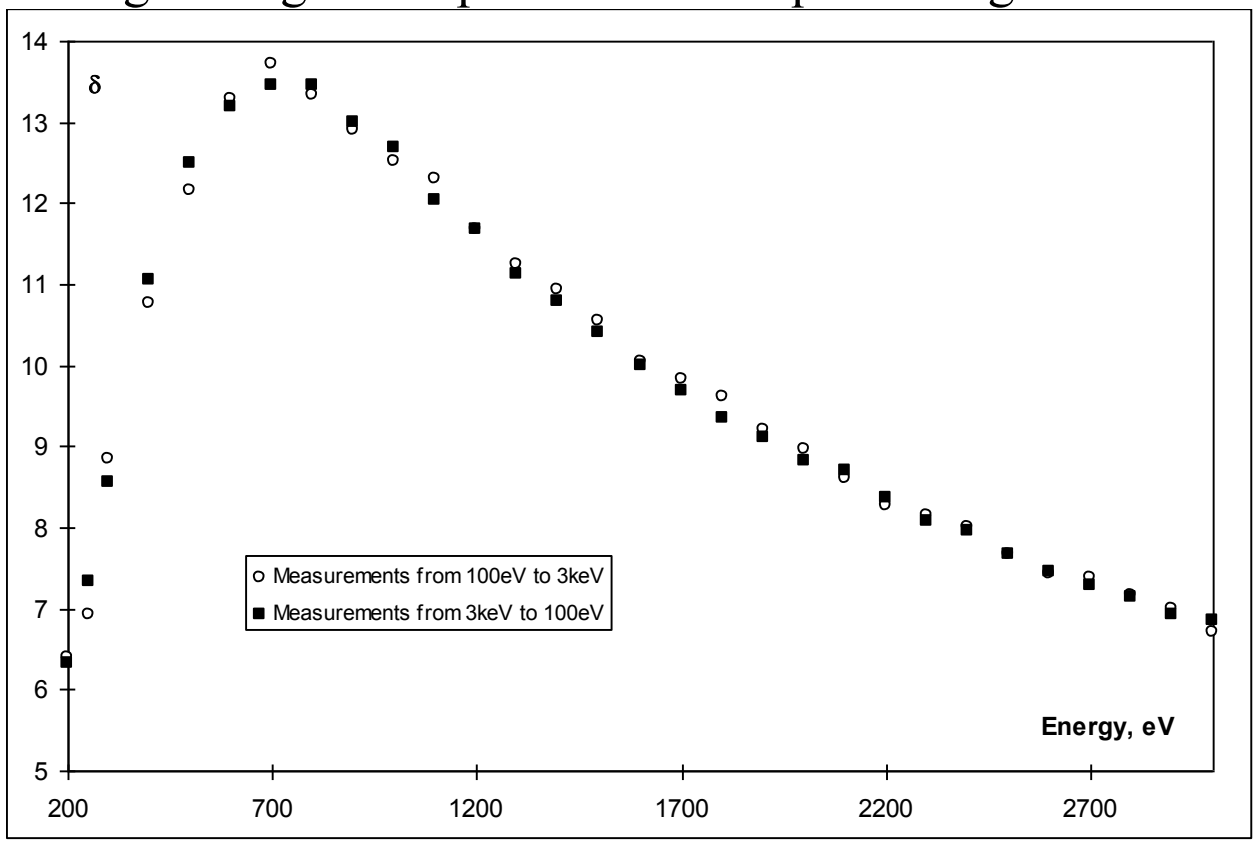

*Fig.23. Comparison of the SEEC results obtained on 100 monolayers of argon depending on direction of measurements..*

It could be interesting to come back again to this remarkable result as influence of direction of the measurements on the SEEC. As it follows from the Fig. 13, the SEEC measurements carried out in direction from low to high energies would lead to incorrect results for the high energy region. This is true for enough great coverages. Similar measurements carried out for low coverages have shown that there are no influence of direction of the measurements on the SEEC. This can be illustrated on the Figure 23 showing that the SEEC measurements carried out for 100 monolayers of argon exhibit about the same results for direction from low to high energies as well as for direction from high to low energies.

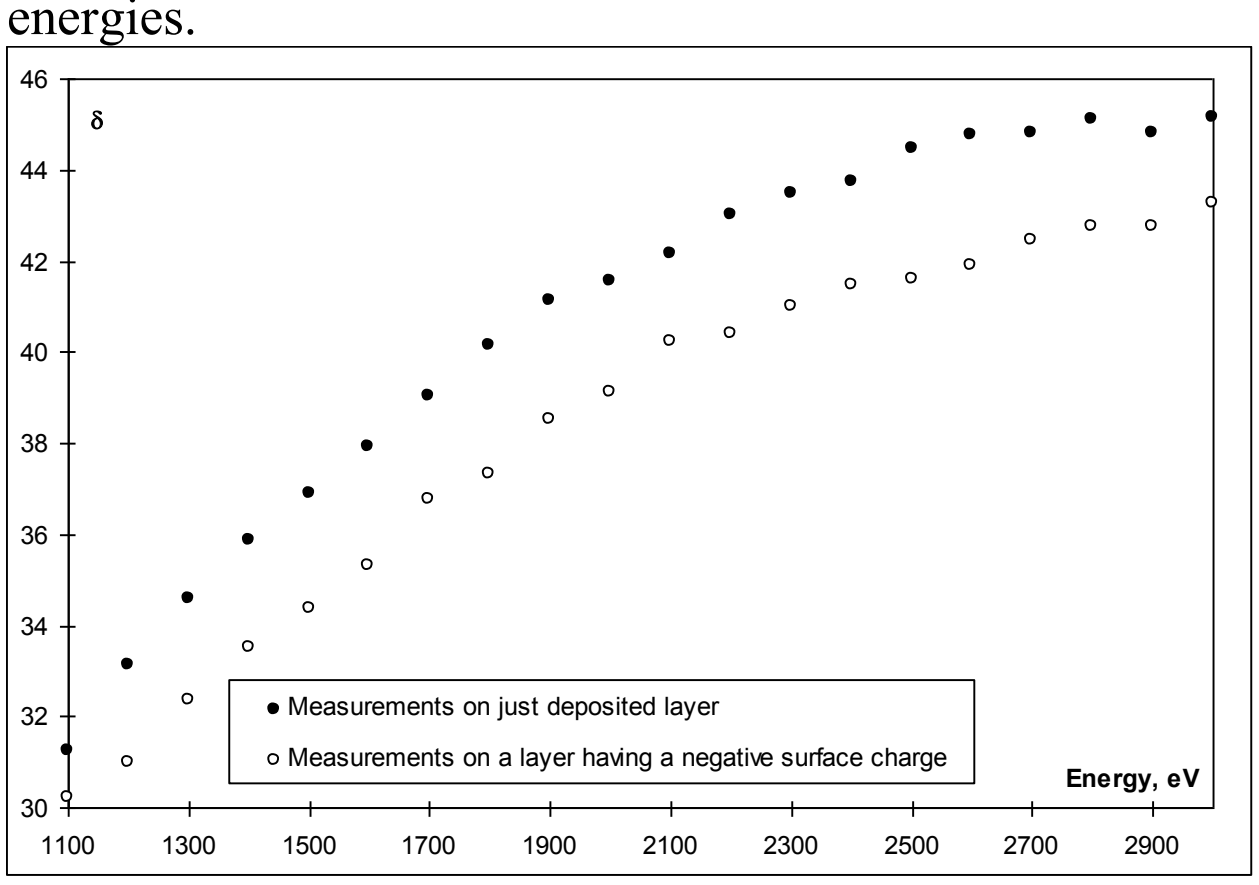

*Fig. 24. Comparison of the SEEC data for a fresh deposited film and for a film having a negative surface charge. Xenon, 500 monolayers.*

Another example illustrating influence of a surface charge concentrated in very thin near surface layer is shown on the Figure 24. This result was obtained during the first cross-over measurements for xenon. Just deposited 500 monolayers of xenon have been bombarded with 9eV incident beam and the current of about $1.5 \cdot 10^{-8}$A. Only one 500μsec pulse was executed during which it was observed that the surface has acquired a negative charge. Just after, the SEEC measurements from 3keV to 1keV have been carried out. Comparing the obtained results with those for fresh 500 monolayers one can see that the presence of the negative charge in near surface layer significantly decreases the secondary electron yield.

These phenomenon related to the thin near surface layer charge could be explained using Young's model[8] stating that the primary beam energy dissipation is approximately constant throughout the range. It means that the near surface layer after a beam impact acquires a space charge but the field across this layer will be negligible. That is, if secondary electrons would originate from this layer, their travel to the surface is not affected. That is why the influence of the direction of the measurements was not observed for low coverages since the penetration depth for high energy primary electrons is much greater of the used coverage. However, when thickness of the film approaches to the penetration depth the primary beam energy dissipation is no longer constant that causes appearance of the field across the film. Secondary electrons originating in the region near the penetration depth will be affected by this field as it was observed in the experiment with negative surface charge and in the experiment with direction of the measurements carried on a thick layer.

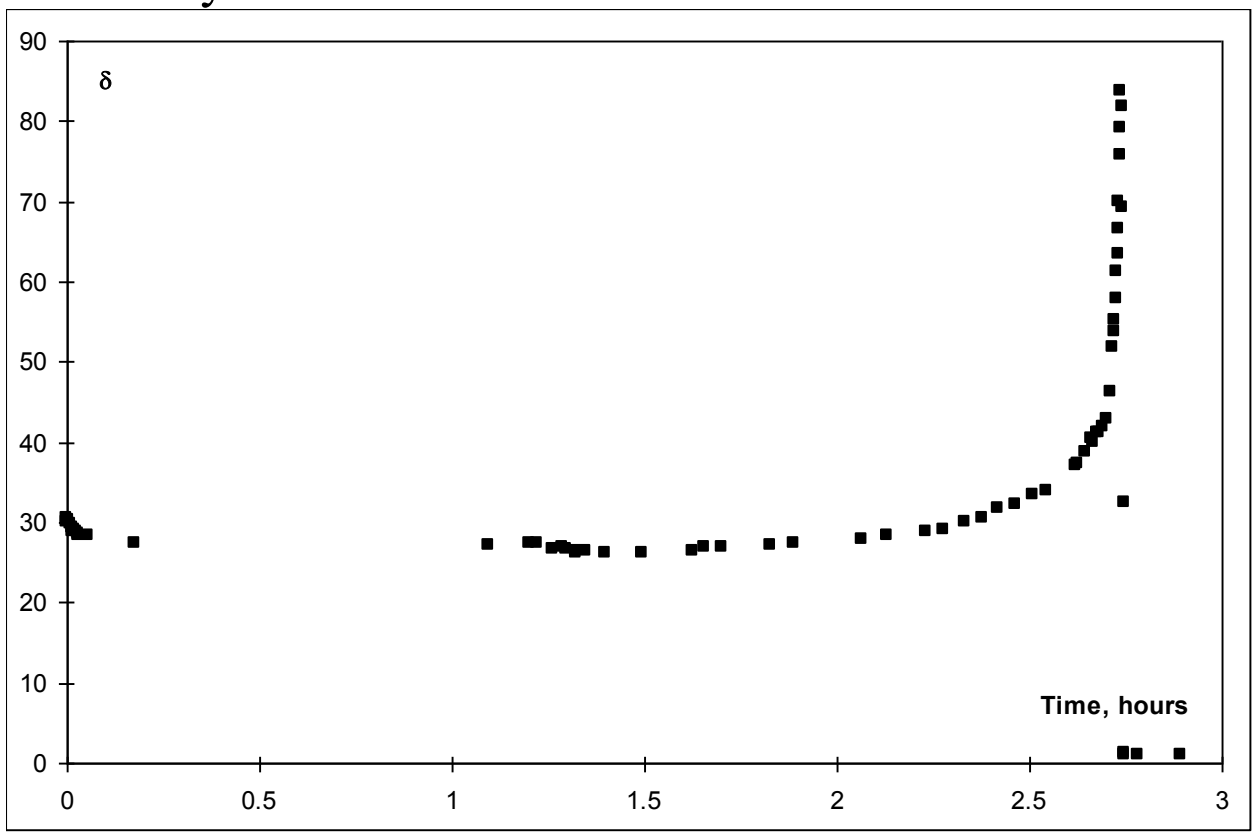

*Fig. 25. Behavior of the SEEC for 2000 monolayers of xenon during warming-up the target. The incident beam energy is 3keV.*

Other interesting results have been observed during the warming up the target with a film of a deposited gas. The Figure 25 shows the SEEC results obtained on 2000 monolayers of xenon for the primary beam energy of 3keV during warming up the target[1]. It can be seen that about 2 minutes before when xenon has completely left the target, the SEEC started its rise to the last maximal measured value of about 90. The beginning of the rise has coincided with appearance of xenon peak in the residual gas analyzer.

[1] One has to note that before doing this experiment the film of deposited xenon was left on the target for many hours and was hence contaminated with residual species. That is why the SEEC values in the beginning of these measurements do not correspond to the ones obtained for a fresh deposited layer.

This experiment seems to be an good illustration of influence of the porous structure of a deposited film on the secondary electron yield. Appearance of the xenon peak in the residual gas analyzer meant that mobility of the xenon atoms on the target has reached its highest value before converting to gaseous state. This increasing with the warming up mobility changed the density of the film making the film more porous. And despite of steadily decreasing film thickness that had to lower the SEEC, influence of the size and number of pores was more dominating mechanism caused finally the SEEC to rise up drastically.

## Acknowledgments.

The authors would like to express their gratitude to Mr. Georges Dominichini for his help during the measurements.

## References

[1] C.Benvenuti, R.S.Calder, CERN-ISR-VA/69-78.
[2] A.J.Dekker, *Solid State Physics*, Vol.6, Academic, New York, (1958).
[3] L.Malter, *Phys. Rev.*, 49, 478, (1936).
[4] L.Malter, *Phys. Rev.*, 50, 48, (1936).
[5] I.Gimpel and O.Richardson, Proc. Roy. Soc. (London) A182, 17 (1943).
[6] J. B. Johnson, K. G. McKay, *Phys. Rev.* Vol.91, p.582 (1953).
[7] J.R.Young, *Phys. Rev.*, 103, 292 (1956).
[8] J.R.Young, J. Appl. Phys. 28, 524 (1957).